\documentclass[sigconf]{acmart}

\AtBeginDocument{%
  }

\copyrightyear{2026}
\acmYear{2026}
\setcopyright{usgovmixed}
\acmConference[CAIS '26]{ACM Conference on AI and Agentic Systems}{May 26--29, 2026}{San Jose, CA, USA}
\acmBooktitle{ACM Conference on AI and Agentic Systems (CAIS '26), May 26--29, 2026, San Jose, CA, USA}
\acmDOI{10.1145/3786335.3813165}
\acmISBN{979-8-4007-2415-2/2026/05}

\usepackage{microtype}
\usepackage{amsfonts}
\usepackage{amsmath}
\usepackage[ruled,linesnumbered]{algorithm2e}
\usepackage{textcomp}
\usepackage{multirow}

\definecolor{myblue}{RGB}{63, 90, 126}
\definecolor{mygray}{RGB}{228, 244, 247}

\newcounter{obsnum}
\newcommand{\observe}[1]{%
  \par\medskip
  \refstepcounter{obsnum}%
  \noindent
  {\setlength{\fboxsep}{4pt}%
  \fcolorbox{myblue}{mygray}{%
    \begin{minipage}{\dimexpr\linewidth-2\fboxsep-3\fboxrule}%
      \vspace{0.5pt}%
      {\bfseries Observation \theobsnum:} #1%
      \vspace{0.5pt}%
    \end{minipage}%
  }}%
  \par\medskip
}
\begin{document}

\title{Understanding and Improving Communication Performance in Multi-node LLM Inference}


\author{Prajwal Singhania}
\orcid{0000-0003-4277-1287}
\affiliation{%
  \institution{Department of Computer Science, University of Maryland}
  \city{College Park}
  \state{MD}
  \country{USA}
}
\email{prajwal@umd.edu}

\author{Siddharth Singh}
\orcid{0000-0002-2756-4290}
\affiliation{%
  \institution{Department of Computer Science, University of Maryland}
  \city{College Park}
  \state{MD}
  \country{USA}
}
\email{siddharth9820@gmail.com}

\author{Lannie Dalton Hough}
\orcid{0009-0000-1589-6530}
\affiliation{%
  \institution{Department of Computer Science, University of Maryland}
  \city{College Park}
  \state{MD}
  \country{USA}
}
\email{ldhough@umd.edu}

\author{Akarsh Srivastava}
\orcid{0009-0000-8404-0664}
\affiliation{%
  \institution{Department of Computer Science, University of Maryland}
  \city{College Park}
  \state{MD}
  \country{USA}
}
\email{akarsh@umd.edu}

\author{Harshitha Menon}
\orcid{0000-0003-4707-9580}
\affiliation{%
  \institution{Lawrence Livermore National Laboratory}
  \city{Livermore}
  \state{CA}
  \country{USA}
}
\email{harshitha@llnl.gov}

\author{Charles Fredrick Jekel}
\orcid{0000-0003-0339-0668}
\affiliation{%
  \institution{Lawrence Livermore National Laboratory}
  \city{Livermore}
  \state{CA}
  \country{USA}
}
\email{jekel1@llnl.gov}

\author{Abhinav Bhatele}
\orcid{0000-0003-3069-3701}
\affiliation{%
  \institution{Department of Computer Science, University of Maryland}
  \city{College Park}
  \state{MD}
  \country{USA}
}
\email{bhatele@cs.umd.edu}

\renewcommand{\shortauthors}{Singhania et al.}

\begin{abstract}
As large language models (LLMs) continue to grow in size, distributed inference
has become increasingly important. Model-parallel strategies must now
efficiently scale not only across multiple GPUs but also across multiple nodes.
In this work, we present a detailed performance study of multi-node distributed
inference using LLMs on GPU-based supercomputers. We conduct experiments with
several state-of-the-art inference engines alongside YALIS, a research-oriented
prototype engine designed for controlled experimentation. We analyze the
strong-scaling behavior of different model-parallel schemes and identify key
bottlenecks.
Because all-reduce operations are a common performance bottleneck, we develop
NVRAR, a hierarchical all-reduce algorithm based on recursive doubling with
NVSHMEM. NVRAR achieves up to 1.9$\times$--3.6$\times$ lower latency than NCCL for message
sizes between 128\,KB and 2\,MB on HPE Slingshot and InfiniBand interconnects.
Integrated into YALIS, NVRAR achieves up to a 1.72$\times$ reduction in end-to-end batch
latency for the Llama 3.1 405B model in multi-node decode-heavy workloads using
tensor parallelism.


\end{abstract}

\ccsdesc[500]{Computing methodologies~Neural networks}
\ccsdesc[500]{Computing methodologies~Distributed algorithms}

\keywords{Machine Learning, Distributed LLM Inference, Communication}

\maketitle

\section{Introduction}
\label{sec:intro}

As large language models (LLMs) grow in size and adoption, inference
costs are rising rapidly~\citep{maslej2025artificialintelligenceindexreport,
bick2024rapid, iea_energy_ai_2025}. Practitioners increasingly rely on larger
models, longer sequences, and compute-intensive reasoning to improve output
quality~\citep{wei2023chainofthoughtpromptingelicitsreasoning,
openai2024openaio1card, snell2024scalingllmtesttimecompute}. Improving inference
performance is therefore critical for reducing energy consumption and
operational costs.

With increasing LLM sizes, memory footprints often exceed the capacity of a
single GPU, requiring parallel execution across multiple devices. On most
clusters, a single node, typically with four to eight GPUs, is insufficient to host
large models such as Llama 3.1 405B~\citep{grattafiori2024llama3herdmodels}.
To enable inference using such models, inference engines use \emph{model
parallelism}~\citep{megatronlm, huang2019gpipe} schemes, which partition the
model parameters across GPUs. While model parallelism for inference 
within a single node has been extensively studied and
optimized~\citep{aminabadi2022deepspeedinferenceenablingefficient,li2024flashcommunicationreducingtensor},
inference in \emph{multi-node} settings is comparatively under-explored. In
this paper, we systematically study, comparatively evaluate, and optimize
multi-node inference workloads.

Multi-node inference introduces challenges such as higher inter-node
latencies compared to faster within-node NVLink connections.
As a result, parallelization strategies that perform well within a node can
experience substantial degradation in multi-node settings due to increased
communication overheads.
Moreover, the optimal choice of parallelism strategy often depends on the
specific inference workload characteristics and the efficiency of underlying
communication libraries. Consequently, it remains unclear
which model parallelism schemes are best suited for multi-node inference and how
to optimize them further. 

In this work, we address the following research questions:
\begin{itemize}
\item How do different model parallel schemes (tensor and hybrid parallelism) scale across
multiple nodes in distributed environments for specific inference workloads?
\item What performance bottlenecks arise in these model parallel
schemes under the workloads studied?
\item Can we optimize collective communication, which appears as a common performance bottleneck
in multi-node inference?
\end{itemize}

To investigate the questions above, we study the performance of two popular
model-parallel schemes: tensor parallelism (TP) and hybrid tensor-pipeline
parallelism (HP), in multi-node settings. We evaluate two state-of-the-art
inference engines, vLLM \citep{woosuk2023vllm} and SGLang
\citep{zheng2024sglang}, alongside YALIS, a research-oriented inference engine
we develop to facilitate controlled experiments in multi-node HPC environments. 
We evaluate the performance of these engines on batched inference workloads, and
study the scaling behavior of both model-parallel schemes. We identify
bottlenecks via detailed performance breakdowns for each of the schemes.

Based on the results of our performance study, we find that workloads that
perform better with TP suffer from significant communication overheads from 
the all-reduce operations. To address this, we propose \emph{NVRAR}: a
hierarchical all-reduce implementation built using
NVSHMEM~\citep{nvshmem}, and optimized for message sizes occurring in inference
workloads. We evaluate NVRAR against NCCL's all-reduce~\citep{nccl} on
multiple HPC interconnects, and observe up to 1.9$\times$ better performance
on HPE Slingshot-11 and 3.6$\times$ on InfiniBand networks in the 256\,KB to 2\,MB
message size range. Integrating NVRAR into YALIS and vLLM yields up to a 1.72$\times$
improvement in multi-node inference performance for the Llama 3.1 405B model in
decode-heavy regimes.

The main contributions of this work are as follows:
\begin{itemize}
    \item We systematically study the performance of model-parallel inference
    schemes in multi-node settings, producing detailed performance breakdowns.
    To facilitate this study, we develop
    YALIS\footnote{\url{https://github.com/axonn-ai/yalis}}, an inference engine
    designed for easier experimentation in multi-node HPC environments.
    \item Based on our performance analysis, we characterize how
    tensor parallelism compares to pipeline parallelism for different
    multi-node workloads and across inference phases. We identify bottlenecks in both
    parallelism schemes.
    \item To address the communication bottleneck
    in multi-node TP inference, we develop
    NVRAR\footnote{\url{https://github.com/hpcgroup/nvrar}}, a custom all-reduce
    implementation optimized for the small-message regime characteristic of
    decode-heavy workloads.
    NVRAR delivers up to 1.72$\times$ faster
    multi-node TP inference for Llama 3.1 405B. 
\end{itemize}

\section{Background}
\label{sec:bg}
This section provides background on LLM inference, model parallelization
strategies, and the communication primitives they use.

LLM inference consists of two phases: \emph{prefill} and \emph{decode}. In
prefill, the model processes all prompt tokens in parallel to generate the
first output token and is typically compute-bound due to large matrix
multiplications. In decode, it generates subsequent tokens
sequentially, becoming memory-bandwidth-bound because of smaller matrix
multiplications and frequent parameter/KV-cache accesses.

\subsection{Model Parallelism for Inference}
LLMs that exceed the memory capacity of a single GPU require distributing model
parameters and computations across multiple GPUs. This is broadly referred to as
\emph{model parallelism}, which can be implemented in several ways.
In \emph{pipeline parallelism (PP)}, contiguous groups of layers are assigned
to $P$ processing units (pipeline stages), forming a sequential dependency
chain with point-to-point communications. It achieves high utilization by
splitting a batch of prompts into pipelined micro-batches.
In \emph{tensor parallelism (TP)}, the computation of each layer is partitioned
across GPUs by splitting the underlying matrix multiplications. TP has no
sequential dependency between GPUs, but aggregation of partial results incurs high
communication overheads due to per-layer all-reduce operations. 

\subsection{Algorithms for All-reduce}
NCCL~\citep{nccl} is the default communication library for AI workloads on
NVIDIA GPUs and primarily implements two all-reduce algorithms: \emph{Ring} and
\emph{Tree}~\citep{hu2025demystifying}. Other variants, such as \emph{CollNet},
depend on specialized DGX hardware and are out of scope for this study. We
model the performance of Ring and Tree all-reduce using the $\alpha$-$\beta$
communication model~\citep{hockney1994communication}. Consider a system with $N$
nodes, each containing $G$ GPUs. The inter-node network has latency
$\alpha_{\mathrm{inter}}$ and bandwidth $\beta_{\mathrm{inter}}$, while the
intra-node interconnect has latency $\alpha_{\mathrm{intra}}$ and
bandwidth $\beta_{\mathrm{intra}}$, where $\alpha_{\mathrm{intra}} <
\alpha_{\mathrm{inter}}$ and $\beta_{\mathrm{intra}} > \beta_{\mathrm{inter}}$. Let
$M$ denote the input message of size $|M|$ bytes.

\vspace{+0.08in}
\noindent \textbf{Ring all-reduce.} NCCL's Ring all-reduce performs a
reduce-scatter followed by an all-gather over a flat ring topology where all
links are active each step. Inter-node links dominate the cost, and the
communication time is modeled as:
\begin{align}
  T_{\mathrm{ring}} = 2(NG-1)\alpha_{\mathrm{inter}} + 2\frac{NG-1}{NG}\left( \frac{|M|}{\beta_{\mathrm{inter}}}\right)
  \label{eq:ring_allreduce}
\end{align}
%
%
%
\noindent \textbf{Tree all-reduce.} The Tree all-reduce performs a reduction
followed by a broadcast using a double binary tree
topology~\citep{nccl_doubletree} for inter-node communication and a simple
intra-node chain. The communication time is modeled as:
\begin{align}
  T_{\mathrm{tree}} \approx& 2 (G - 1)\, \alpha_{\mathrm{intra}} + 2 \log_2 (N)\, \alpha_{\mathrm{inter}} 
    + 2 \frac{N - 1}{N}\left( \frac{|M|}{\beta_{\mathrm{inter}}} \right)
   \label{eq:tree_allreduce}
\end{align}
For simplicity, we consider only the inter-node bandwidth term in the above
expression ($\beta_{\mathrm{intra}} \gg \beta_{\mathrm{inter}}$). 

\noindent \textbf{NVSHMEM.} An OpenSHMEM-based~\citep{chapman2010openshmem}
communication library from NVIDIA, providing host and device APIs for one-sided
put/get and collective operations over NVLink, Slingshot, and InfiniBand. It
enables implementation of GPU-initiated communication kernels.

\section{Studying Performance of Multi-node Inference}
\label{sec:perfstudy}
This section presents a performance study of multi-node LLM inference. Our
objective is to evaluate different model-parallelism schemes, understand
their scaling behavior, and identify bottlenecks for batched inference workloads.
We first introduce YALIS, a prototype inference engine built for controlled performance studies.
We then detail our benchmarking methodology and present our experimental results,
discussing the performance of YALIS and existing state-of-the-art inference engines
in multi-node settings.

\subsection{YALIS: Yet Another LLM Inference System}
YALIS is an open-source inference engine built as a research vehicle to study
multi-node LLM inference. It is intended to be performant, easy to instrument,
and more amenable to Slurm-based environments. These properties allow for
detailed analysis of inference performance on HPC systems. Its
design is centered around three key components: (1) a unified model definition
layer, adapted from LitGPT~\citep{litgpt-2023}, providing compatibility with a
wide range of model architectures; (2) an execution layer utilizing Torch
Compile~\citep{torchcompile} for kernel fusion and optimization, and CUDA
Graphs~\citep{cudagraphs} for minimal kernel-launch overheads; and (3) tensor
model parallelism implemented via AxoNN~\citep{singh:ipdps2022,
singh:sc2024,singh:arxiv2024}, both within and across nodes. 

\begin{figure*}[t]
  \centering
  \includegraphics[width=0.33\textwidth]{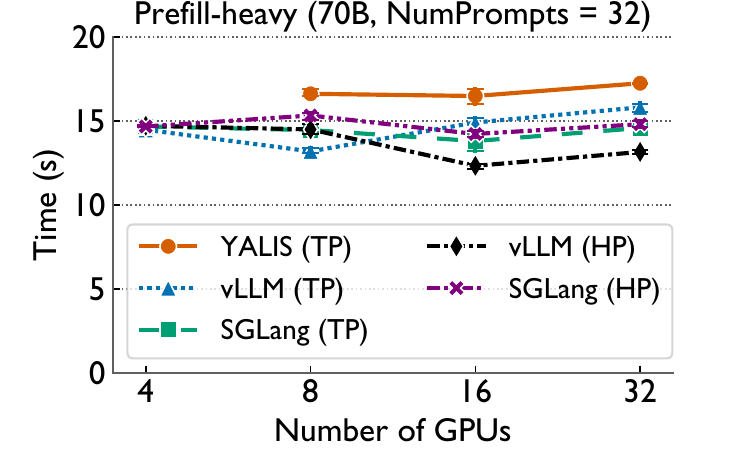}
  \includegraphics[width=0.33\textwidth]{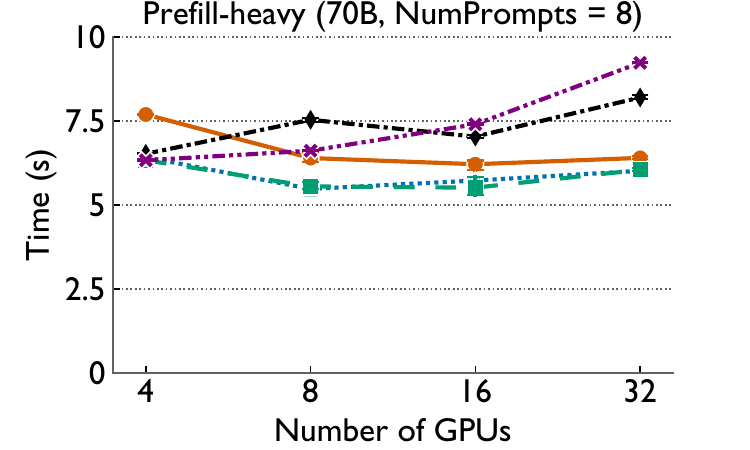}
  \includegraphics[width=0.33\textwidth]{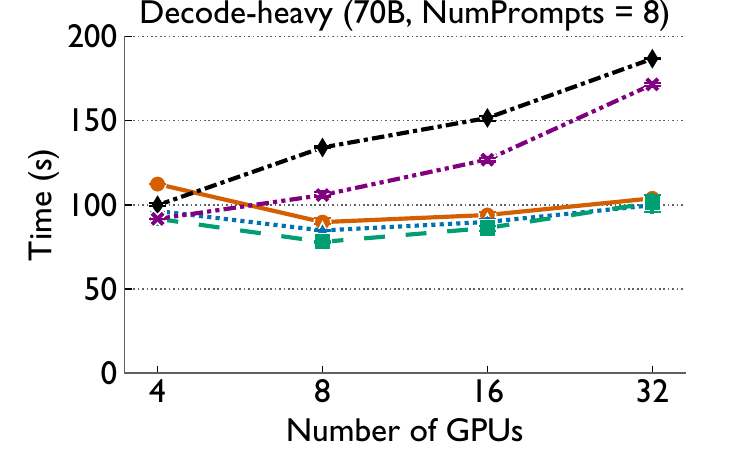}

  \caption{Strong scaling performance of different inference engines on
  Perlmutter for Llama 3.1 70B Instruct. The Y-axis denotes the end-to-end latency per batch in
  seconds, and the X-axis denotes the number of GPUs.}
  \label{fig:strong_scaling_70b}
\end{figure*}

\begin{figure*}[t]
  \centering
  \includegraphics[width=0.33\textwidth]{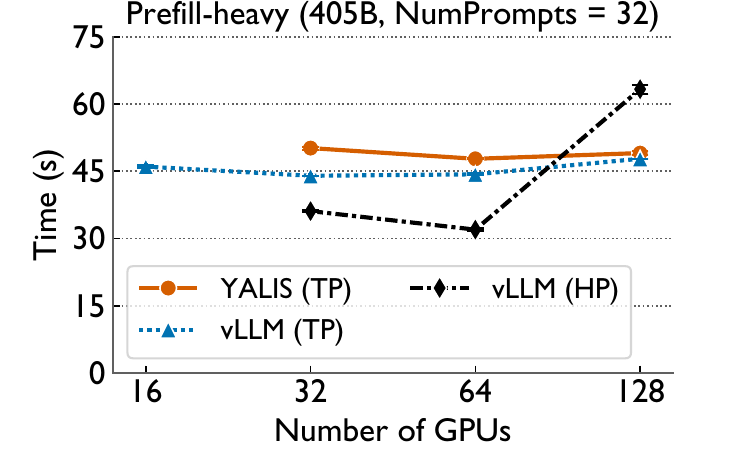}
  \includegraphics[width=0.33\textwidth]{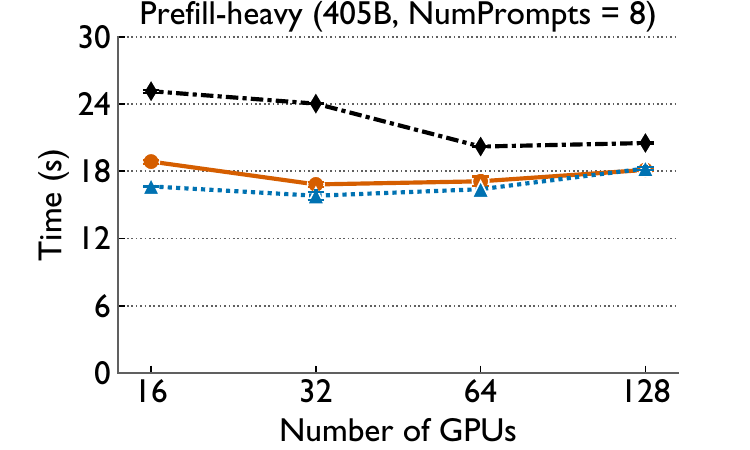}
  \includegraphics[width=0.33\textwidth]{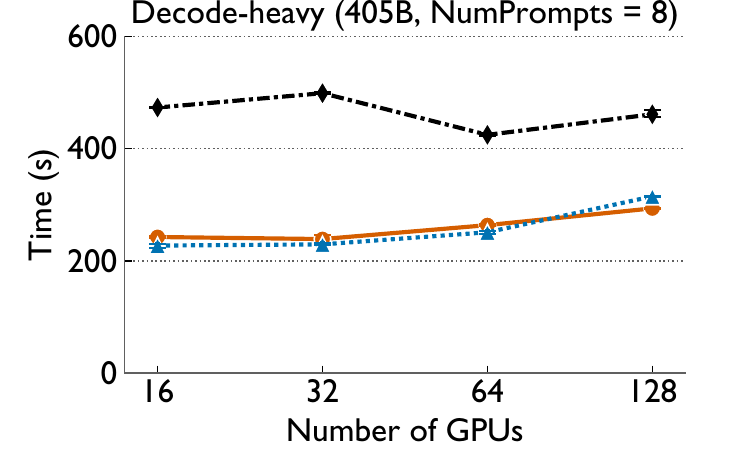}

  \caption{Strong scaling performance of different inference engines on
  Perlmutter for Llama 3.1 405B Instruct. The Y-axis denotes the end-to-end latency per batch in
  seconds and the X-axis denotes the number of GPUs.}
  \label{fig:strong_scaling_405b}
\end{figure*}

\subsection{Benchmarking Methodology}

\begin{table}[h]
  \caption{Details of the HPC systems used in our experiments.}
  \centering
  \resizebox{\columnwidth}{!}{
  \begin{tabular}{llll}
    \toprule
    \textbf{System} & \textbf{GPU} & \textbf{GPUs/Node} & \textbf{Interconnect} \\
    \toprule
    \multirow{2}{*}{Perlmutter} & \multirow{2}{*}{A100 (40/80\,GB)} & \multirow{2}{*}{4} & Intra-Node: 3$^{\text{rd}}$ gen NVLink \\
                                &                                    &                    & Inter-Node: Slingshot-11 \\
    \midrule
    Vista & GH200 (96\,GB) & 1 & Inter-Node: InfiniBand \\
    \bottomrule
  \end{tabular}
  }
  \label{tab:machines}
  \vspace{-0.2cm}
\end{table}
\noindent \textbf{Hardware and Models.} Our scaling experiments use the Perlmutter
system~\citep{perlmutter} (Table~\ref{tab:machines}) (80\,GB nodes unless
otherwise specified). We evaluate two dense LLMs --- Llama 3.1 70B (Instruct)
and 405B (Instruct)~\citep{grattafiori2024llama3herdmodels} --- in bf16
precision. While we evaluate a single model family, its architecture and
compute, memory, and communication patterns are representative of other dense
LLMs.

\begin{table}[h]
  \caption{Details of workloads evaluated in our experiments.}
  \label{tab:workloads}
  \centering
  \resizebox{\columnwidth}{!}{
  \begin{tabular}{llll}
    \toprule
    \textbf{Workload} & \textbf{Prompt Length} & \textbf{Decode Length} & \textbf{NumPrompts (\#P)} \\
    \midrule
    Prefill-heavy & 2363 & 128 & 8, 32 \\
    Decode-heavy & 1426 & 3072 & 8, 32 \\
    \bottomrule
  \end{tabular}
  }
  \vspace{-0.2cm}
\end{table}
\noindent \textbf{Workloads and Metrics.} Table~\ref{tab:workloads} lists the workload
configurations used in our experiments. We define \emph{NumPrompts} (\#P) as the
number of prompts provided to the inference engine in a single user batch. For
brevity, we present a subset of results in the main text, with additional
results in Appendix~\ref{sec:appendix_perfstudy}. 

In \emph{batched inference} workloads, the engine processes
one batch of prompts to completion before we submit the next batch. This mirrors
real-world settings such as synchronous GRPO~\citep{shao2024deepseekmath,
guo2025deepseek} and helps isolate GPU execution performance from scheduler
effects. We report the total time-to-completion for a single batch of prompts.

In our strong scaling experiments (fixed workload across GPUs), the 70B model is
scaled from four GPUs (single node) to 32 GPUs (eight nodes), and the 405B model
is scaled from 16 GPUs (four nodes) to 128 GPUs (32 nodes). Each run includes
two warm-up and up to three timed generations. We repeat each run
three times and report average performance. For performance breakdowns, we use one run
with two warm-up and one profiled generation.

\begin{table}[h]
  \caption{Parallelism schemes and inference engines.}
  \label{tab:frameworks}
  \centering
  \resizebox{\columnwidth}{!}{
  \begin{tabular}{llll}
    \toprule
    \textbf{Parallelism} & \textbf{Intra-Node} & \textbf{Inter-Node} & \textbf{Engines}\\
    \toprule
    \multirow{2}{*}{Tensor Parallelism (TP)} & \multirow{2}{*}{TP} & \multirow{2}{*}{TP} & YALIS, vLLM V1 (v0.11.0),\\
    &                                    &                    & SGLang (v0.5.1) \\
    \midrule
    \multirow{2}{*}{Hybrid Parallelism (HP)} & \multirow{2}{*}{TP} & \multirow{2}{*}{PP} & vLLM V0 (v0.10.0),\\
    &                                    &                    & SGLang (v0.5.1) \\
    \bottomrule
  \end{tabular}
  }
  \vspace{-0.2cm}
\end{table}
\noindent \textbf{Software Stack and Parallelism Schemes.}
Table~\ref{tab:frameworks} lists the inference engines we use in our experiments
for different parallelism schemes. We use PyTorch 2.8~\citep{paszke2019pytorch}
and CUDA 12.9 for all experiments. For vLLM, we use the V0 engine for HP because
we observed persistent hangs with the V1 engine when using Ray-based PP on
Slurm-based systems~\citep{ray_issue58426}. For performance breakdowns, we use
Nsight Systems~\citep{nsys} to collect traces and
Pipit~\citep{bhatele:2023pipit} to analyze them.

\subsection{Scaling Multi-node LLM Inference}
Figures~\ref{fig:strong_scaling_70b} and~\ref{fig:strong_scaling_405b} report the
time-to-completion for a batch of prompts across all engines for Llama
3.1 70B and 405B on Perlmutter, respectively. From left to right, the
workloads transition from compute-bound to increasingly memory-bound regimes. 

We first observe that YALIS (orange line) achieves comparable performance to
state-of-the-art engines, particularly for memory-bound workloads. For the
70B model, YALIS is within 5--16\% of vLLM (TP) at eight GPUs and beyond. For the
405B model, it is within 8\% for all GPU counts. The only noticeable
deviation occurs for the 70B model's prefill-heavy workload at 16 GPUs.
Crucially, YALIS exhibits scaling trends consistent with other engines,
validating its suitability as a research vehicle for studying multi-node LLM
inference. The missing data points correspond to OOM errors.

Across all models and engines, both TP and HP exhibit poor strong scaling, where
the time to solution does not scale inversely with GPU count. Focusing on the
70B model (Figure~\ref{fig:strong_scaling_70b}), vLLM (TP) (green line)
latencies decrease from four GPUs (single node) to eight GPUs (two nodes), with
more noticeable improvements for the decode-heavy workload (right-most plot).
However, after 16 GPUs, the latency remains almost constant or increases with
GPU count. This trend is consistent for TP, across all engines and models.

When using HP, we observe a different trend. In the prefill-heavy regime for
the 70B model (Figure~\ref{fig:strong_scaling_70b}), vLLM (HP) (black line)
latencies remain nearly constant with a smaller number of prompts (middle plot),
but decrease (up to 16 GPUs) with a larger number of prompts (left plot). For the
405B model, latencies decrease initially for both small and large numbers of
prompts, before increasing or flattening out. SGLang (HP) (pink line) exhibits a 
similar trend. For decode-heavy workloads, however, HP latencies increase
significantly with increasing GPU count for both vLLM and SGLang.

Comparing the two schemes, HP outperforms TP for the most compute-bound and
prefill-heavy workload (Figure~\ref{fig:strong_scaling_70b}, left), but TP starts
to outperform HP as workloads become more memory-bound and decode-heavy. This holds across
engines (Figure~\ref{fig:strong_scaling_70b}, middle and right). Similar trends are
observed for the 405B model (Figure~\ref{fig:strong_scaling_405b}).
\observe{ For the workloads studied, TP and HP do not scale ideally. HP is 
advantageous in compute-bound regimes, whereas TP is better for
memory-bound and decode-heavy cases.}
\subsection{Identifying Performance Bottlenecks}
\begin{figure}[h]
  \centering
  \includegraphics[width=\columnwidth]{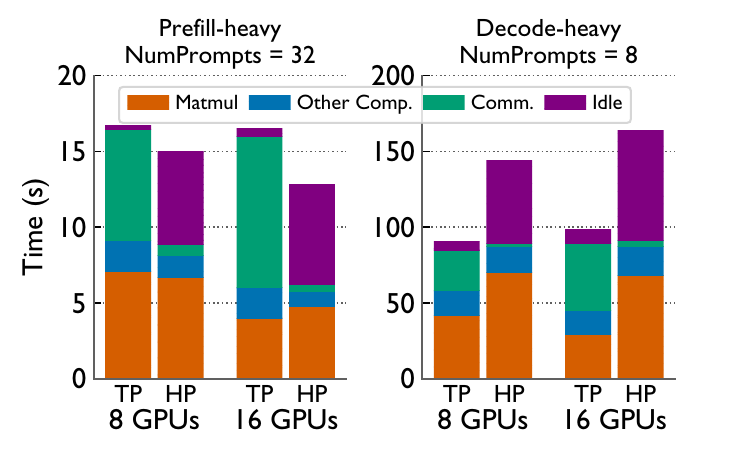}
  \caption{Performance breakdown of TP (using YALIS) and HP (using vLLM) for
  the prefill-heavy and decode-heavy workloads on Perlmutter for the 70B Llama
  model.}
  \label{fig:breakdown_pp_v_yalis}
  \vspace{-0.25cm}
\end{figure}

To better understand the scaling behaviors of TP and HP across prefill- and
decode-heavy workloads, we analyze the performance breakdowns of YALIS (TP) and
vLLM (HP) on eight and 16 GPUs for the 70B model
(Figure~\ref{fig:breakdown_pp_v_yalis}). We decompose the total time into four
components: \emph{Matmul} (time spent in matrix multiplications), \emph{Other
Comp.} (time spent in other computations), \emph{Comm.} (time spent in
communication), and \emph{Idle} (per-GPU idle time). We report per-GPU
breakdowns rather than aggregating across the critical path.

\begin{table}[h]
  \caption{Synthetic benchmarks modeling Prefill-GEMM ($M$=32768, $N$=8192,
  $K$=57344) and Decode-GEMM ($M$=32, $N$=8192, $K$=57344) matrix
  multiplications in the MLP layer of the 70B Llama model. $M$/$2$ corresponds
  to HP micro-batching and $K$/$2$ corresponds to TP.}
  \centering
  \resizebox{\columnwidth}{!}{
  \begin{tabular}{lccc}
    \toprule
    \textbf{Workload} & \textbf{Baseline ($M$,$N$,$K$)} & \textbf{HP ($M$/$2$,$N$,$K$)} & \textbf{TP ($M$,$N$,$K$/$2$)} \\
    \toprule
    Prefill-GEMM & 108.033 ms & 53.824 ms & 53.896 ms \\
    Decode-GEMM & 0.614 ms & 0.574 ms & 0.359 ms \\
    \bottomrule
    \end{tabular}
  }
  \label{tab:gemm_benchmark}
  \vspace{-0.25cm}
\end{table}

For the prefill-heavy workload (Figure~\ref{fig:breakdown_pp_v_yalis}, left),
both YALIS (TP) and vLLM (HP) reduce computation time going from eight to 16
GPUs, with vLLM (HP) achieving lower overall latency due to reduced
communication overhead. However, vLLM (HP) exhibits unexpectedly high GPU
idle time. We hypothesize that this behavior is due to pipeline bubbles in
batched inference caused by imbalanced prefill and decode stage times.
Continuous batching~\citep{yu2022orca} can mitigate this, but we 
leave detailed investigation to future work.

For decode-heavy workloads, HP fails to reduce the time spent in matrix
multiplications, unlike TP. This partially explains why, despite lower
communication costs, HP does not scale as well for such workloads. To
isolate this behavior, we run a synthetic GEMM benchmark using two
representative matrix sizes: Prefill-GEMM ($M$=32768, $N$=8192, $K$=57344) and
Decode-GEMM ($M$=32, $N$=8192, $K$=57344). The former models large-$M$ prefill
matmuls (batch size $\times$ prompt length), while the latter models small-$M$
decode matmuls (batch size $\times$ 1). Table~\ref{tab:gemm_benchmark} reports
the runtime of both when either $M$ is halved (micro-batching in the PP phase
of HP) or $K$ is halved (TP). For Prefill-GEMM, halving either $M$ or $K$ nearly
halves the runtime. For Decode-GEMM, however, halving $K$ reduces the
runtime substantially, whereas halving $M$ yields only
a marginal reduction. This behavior arises due to tiling in GEMM kernels, where
decreasing $M$ below the tile size yields no speedup. While TP
outperforms HP for these workloads, it still incurs significant communication
overhead. Figure~\ref{fig:breakdown_pp_v_yalis} (right) highlights that the
communication time in YALIS (TP) increases by $\sim$1.6$\times$, offsetting the
gains from reduced computation time when going from eight to 16 GPUs.
\observe{ For prefill-heavy workloads, both TP and PP reduce computation time,
with PP achieving lower overall latency due to its reduced communication
overhead. For decode-heavy workloads, PP does not reduce matrix multiplication
time, while TP suffers from significant communication overhead.}

\subsection{Communication Issues in Tensor Parallelism}
\begin{figure}[h]
  \centering
  \includegraphics[width=\columnwidth]{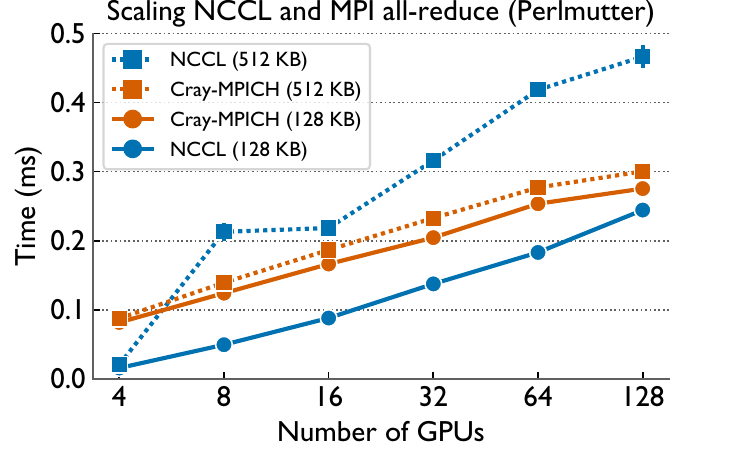}
  \caption{Scaling performance of NCCL and MPI all-reduce for a range of message sizes on Perlmutter.}
  \label{fig:nccl_mpi_scaling}
\end{figure}

The primary communication collective in TP is all-reduce. In the
decode-heavy regime, it is dominated by small messages
of size $B \times H$, where $B$ is the batch size and $H$ the hidden dimension.
For the 70B model with $B$=8 and $H$=8192, this message size is 128\,KB. 
Across our workloads, message sizes range from 128\,KB to 1\,MB.

To further analyze communication bottlenecks, we benchmark NCCL all-reduce
against GPU-aware Cray-MPICH on Perlmutter (40\,GB nodes), focusing on small
messages. We run the OSU benchmark~\citep{osu-5.8} and
\texttt{nccl-tests}~\citep{nccl-tests}, and report average all-reduce time over
10 runs (200 warm-up and 10,000 timed iterations) in
Figure~\ref{fig:nccl_mpi_scaling}. We observe that NCCL all-reduce is substantially
faster within a node, but scales poorly across nodes compared to MPI. For
512\,KB--1\,MB messages, NCCL is 1.5--2$\times$ slower than MPI, with latency
growing faster with message size at any given scale. While Cray-MPICH's
implementation is proprietary, the open-source MPICH~\citep{MPICH} library
typically employs a latency-optimal recursive-doubling
algorithm~\citep{thakurimproving2003} for these regimes, which can explain the
performance gap.
\observe{
  For small message sizes, typical in the decode phase, NCCL all-reduce exhibits
  poor scaling across nodes and can at times be slower than MPI.}

\section{Optimized Multi-node All-reduce}
\label{sec:allreduce}
\begin{figure*}[t]
  \centering
  \includegraphics[height=1.5in]{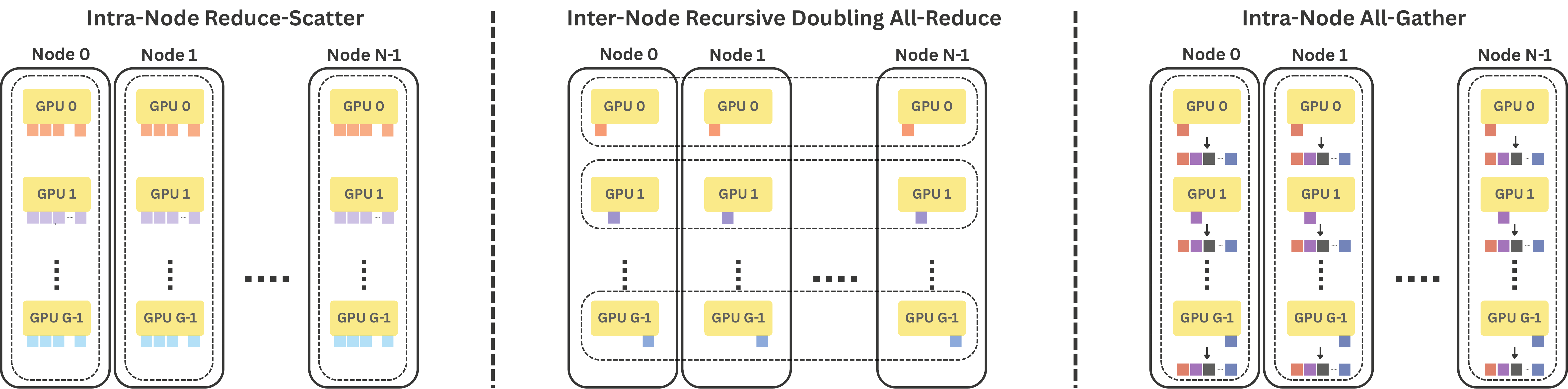}
  \caption{Three-phase NVRAR design: (1) intra-node reduce-scatter, (2) inter-node all-reduce, (3) intra-node all-gather.}
  \label{fig:nvrar_design}
\end{figure*}

We establish in the previous section that TP is generally more performant for
decode-heavy workloads. Next, we focus on optimizing its communication
bottlenecks. One potential approach to address NCCL's all-reduce performance
issues is to use MPI wherever it is faster than NCCL. However, standard MPI
implementations are ill-suited for inference workloads due to the lack of CUDA
Graph support and sub-optimal intra-node NVLink communication. As an alternative
to NCCL and MPI, we develop NVRAR: an NVSHMEM-based hierarchical all-reduce
implementation that uses the recursive-doubling algorithm and is optimized for
small-message inter-node communication.

\begin{algorithm}[t]
\caption{NVRAR}
\label{alg:nvrar}
\small
\KwIn{Message $M$; GPUs/node $G$; number of nodes $N$; chunk size $C_s$;
  sequence number $seq$; GPU rank (within node) $r_g$; node rank $r_n$;
  pre-allocated send/receive buffers $B_{send}, B_{recv}$}
\KwOut{$M$ reduced in-place}
\SetKwProg{Fn}{Function}{:}{}
\Fn{\upshape NVRAR($M, G, N, C_s, seq, r_g, r_n$)}{
  $M^{\prime} \gets \mathrm{REDUCE\mbox{-}SCATTER}_{intra}(M, G)$\;
  $seq \gets seq + 1$\;
  \For{$i = 0$ \KwTo $\log_2 N - 1$}{
    $peer_{i} \gets (r_n \oplus 2^{i}, r_g)$\;
    \textsc{Wait}$(peer_{i},\, \textit{seq})$ \tcp{Synchronize Sequence Number}
  }
  $B_{send}[0] \gets \textsc{PackDataAndSeqNum}(M^{\prime}, \textit{seq})$\;
  $B_{out} \gets \mathrm{RD_{inter}}([B_{send}, B_{recv}], N, C_s, seq, r_g, r_n)$\;
  $M^{\prime} \gets \textsc{UnpackDataAndSeqNum}(B_{out}, seq)$\;
  $M \gets \mathrm{ALL\mbox{-}GATHER}_{intra}(M^{\prime}, G)$\;
}
\BlankLine
\Fn{$\mathrm{RD_{inter}}$($B_{send}, B_{recv}, N, C_s, seq, r_g, r_n$)}{
  $Q \gets \lceil |B_{send}[0]| / C_s \rceil$ \tcp{Number of chunks}
  \For{$\ell = 0$ \KwTo $\log_2 N - 1$}{
    $peer_{\ell} \gets (\,r_n \oplus 2^{\ell},\ r_g\,)$\;
    \For{$q = 0$ \KwTo $Q - 1$}{
      $src_q \gets B_{send}[\ell][q]$;\enspace $dst_q \gets B_{recv}[\ell][q]$\;
      $\mathrm{NON\mbox{-}BLOCKING\mbox{-}PUT}{:}\ src_q \rightarrow peer_{\ell}$'s $dst_q$\;
      Wait until $flag(dst_q) == seq$ \tcp{Wait for data}
      $B_{send}[\ell+1][q] \gets dst_q + src_q$\;
    }
  }
  \Return{$B_{send}[\ell]$}\;
}
\end{algorithm}

\subsection{Three-Phase Hierarchical All-reduce Design}
NVRAR (Algorithm~\ref{alg:nvrar}) has three phases: (1) an intra-node reduce-scatter, (2) an
inter-node recursive-doubling all-reduce, and (3) an intra-node all-gather.
Figure~\ref{fig:nvrar_design} illustrates this design for an
$N$-node system with $G$ GPUs/node. 

\vspace{0.08in}
\noindent \textbf{Reduce-scatter Phase:} In the first phase (Line 2 of
Algorithm~\ref{alg:nvrar}), GPUs within a node perform a local reduce-scatter
operation. For input message $M$ of size $|M|$ bytes, each GPU holds
$\frac{|M|}{G}$ bytes of data reduced within each node at the conclusion of this
phase. We implement this using the
\texttt{nvshmemx\_TYPENAME\_sum\_reducescatter} host-side API, which internally calls NCCL
reduce-scatter. 

\vspace{0.08in}
\noindent \textbf{Inter-Node Recursive-Doubling Phase:} In the second phase
(Line 9 of Algorithm~\ref{alg:nvrar}), corresponding GPUs with the same local ID on each node perform an
all-reduce. Each node is identified by its ID $r_n \in [0,
N-1]$, and each GPU within a node by its local ID $r_g \in [0, G-1]$, so each
GPU is uniquely identified by the pair $(r_n, r_g)$. This phase takes $\log_2 N$
steps. At each step $0 \leq i < \log_2 N$, GPU $(r_n, r_g)$ exchanges data with
its $2^i$-th logical peer, $(r_n \oplus 2^i, r_g)$, where $\oplus$ denotes
bitwise XOR. Thus, GPUs with the same local ID communicate across nodes. 
Upon receiving data, each GPU performs a local reduction with the
received buffer before proceeding to the next step. After all $\log_2 N$ steps,
each GPU holds $\frac{|M|}{G}$ bytes of the globally reduced data. We implement
this phase with a custom NVSHMEM kernel using non-blocking
\texttt{put\_nbi}-based RMA primitives.

\vspace{0.08in}
\noindent \textbf{All-gather Phase:} In the third and final phase (Line 11 of
Algorithm~\ref{alg:nvrar}), the GPUs within a node perform a local all-gather
operation to combine their $\tfrac{|M|}{G}$ fraction of the globally reduced data
into a single tensor. Similar to the reduce-scatter phase, we implement this
using NVSHMEM's host API.
After completion, every GPU holds the full globally reduced tensor, completing the
all-reduce.

\subsection{Performance Optimizations}
The inter-node phase of NVRAR contributes most to the overall all-reduce runtime.
Apart from choosing the algorithmically optimal recursive-doubling approach, we
make three key optimizations for increased efficiency and lower latency: (1)
chunked non-blocking communication, 
(2) fused data-flag payloads for per-step synchronization,
and (3) sequence-number-based global synchronization.

\subsubsection{Chunked Non-Blocking Communication}
Efficient utilization of GPU SMs and concurrent progress of data transfers
and reductions across thread blocks are crucial for all-reduce performance. To
achieve this, we partition the message into disjoint data blocks, processed
independently by $B_s$ thread blocks. Each thread block further subdivides its
data into chunks of size $C_s$ bytes (Lines 15--21 in Algorithm~\ref{alg:nvrar}).
Each chunk is transmitted to the corresponding peer GPU using a non-blocking,
block-cooperative NVSHMEM primitive. Upon receiving a peer's chunk, the thread
block performs a local reduction and advances to the next chunk. 

This design allows different thread blocks to progress through different stages
of the all-reduce concurrently, creating coarse-grained
computation-communication overlap across SMs. Non-blocking communication allows
asynchronous sending and waiting for data, while per-block chunking offers
tunable control over the granularity of network injection. We tune $B_s$ and $C_s$
once for a given message size and node count, as performance is sensitive to
these hyperparameters (Appendix~\ref{sec:appendix_chunk_block}).

To further enable asynchronous progress across inter-node steps, we allocate
per-step send and receive buffers. This allows sending data to the next peer
before the receiving peer has completed its previous step. The extra memory
overhead is negligible for small messages and logarithmic recursion depth.

\subsubsection{Fused Payloads for Step Synchronization}
At each recursive-doubling step, peers need to synchronize to ensure the
completion of remote data receipt before performing local reduction. A
straightforward approach is to use NVSHMEM's explicit signaling primitives ---
\texttt{put\_with\_signal} and \texttt{wait\_until}. However, these
primitives introduce non-trivial latency overheads, particularly on Slingshot
networks. The root cause lies in NVSHMEM's current libfabric implementation,
where \texttt{put\_with\_signal} relies on software fences instead of
Slingshot's hardware fences and message ordering capabilities.

To avoid explicit signaling, we adopt NCCL's low-latency \texttt{LL} protocol
design, fusing data and synchronization flags into a single 8\,B payload (4\,B data +
4\,B flag). This granularity ensures atomic and ordered delivery of each data word and
its flag, on both Slingshot and InfiniBand. Fused payloads also allow
reductions to begin immediately upon receipt (at a warp level), enabling
fine-grained progress and synchronization without extra communication overhead.

\begin{figure*}[t]
  \centering
  \includegraphics[width=0.33\textwidth]{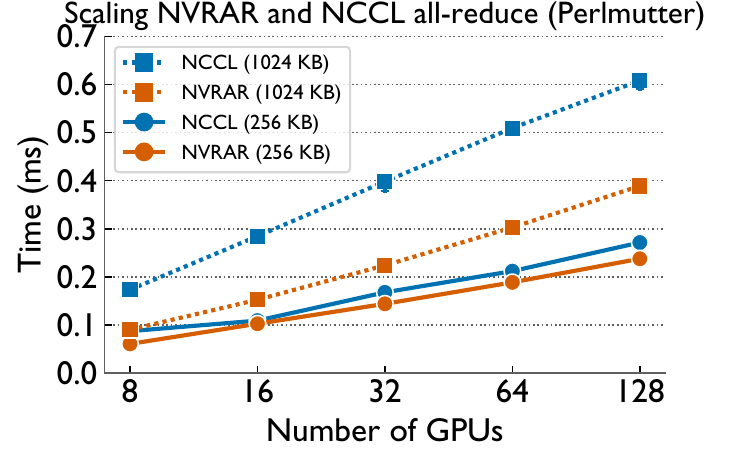}
  \includegraphics[width=0.33\textwidth]{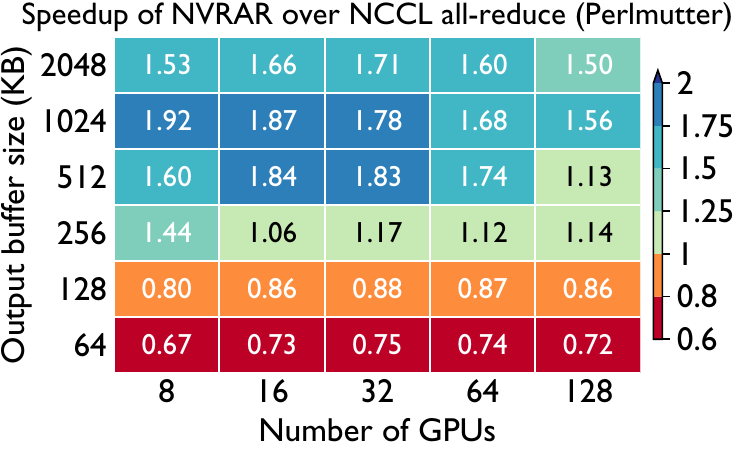}
  \includegraphics[width=0.33\textwidth]{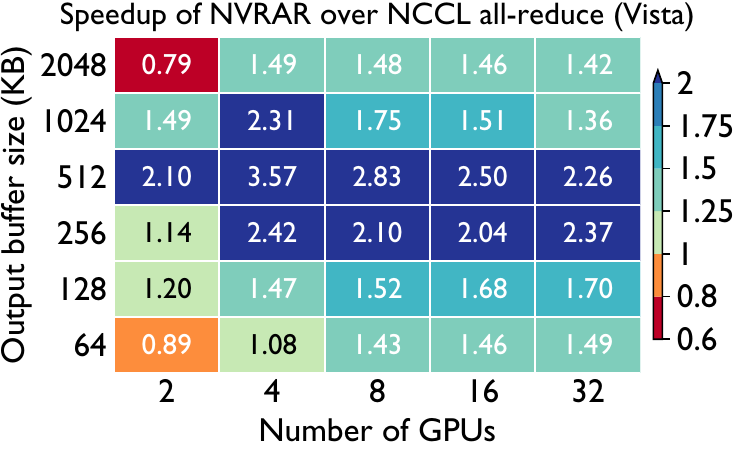}
  \caption{
    Performance comparison of NVRAR and NCCL all-reduce. (Left) Scaling of
    NVRAR and NCCL all-reduce for 256\,KB and 1024\,KB messages across GPU
    counts on Perlmutter. (Middle, Right) Speedup of
    NVRAR over NCCL across message sizes and GPU counts on Perlmutter (A100, Slingshot-11) and
    Vista (GH200, InfiniBand).}
  \label{fig:scaling_nccl_nvrar_pm}
\end{figure*}

\subsubsection{Sequence-Number-Based Global Synchronization}
\label{subsec:seqnum_sync}
When all-reduce operations are issued in succession, previous operations 
must complete before the intermediate buffers can safely be reused. NVSHMEM
provides \texttt{quiet} and \texttt{fence} primitives to achieve this, but they
add significant latency overheads. Instead, we assign each all-reduce operation
a unique sequence number (Line 2 of Algorithm \ref{alg:nvrar}). Each rank
waits for its peers to reach the same sequence number (Lines 4--6 of Algorithm
\ref{alg:nvrar}) before sending data in the inter-node communication phase.

Crucially, each rank synchronizes only with its peers, rather than with all
ranks through a global barrier. This synchronization occurs at the beginning of
the all-reduce operation, rather than at the end, allowing each rank to use the
all-reduced data immediately after its operation completes. Waiting for peer
ranks to finish is deferred until the next all-reduce is issued. We implement
this using NVSHMEM's atomics. 

\subsection{Performance Model for NVRAR}

To analyze NVRAR's performance, we model its communication time using the
$\alpha$-$\beta$ model and notations from Section~\ref{sec:bg}.

\noindent \textbf{Reduce-scatter Phase.} Within a node, NVRAR uses NCCL
reduce-scatter (Ring), modeled as:
\begin{align}
  T_{\mathrm{RS}} = (G-1)\alpha_{\mathrm{intra}} 
  + \frac{G-1}{G}\left( \frac{|M|}{\beta_{\mathrm{intra}}} \right)
\end{align}
\noindent \textbf{Inter-Node Recursive-Doubling Phase.} The inter-node phase
proceeds in $\log_2 (N)$ steps across $N$ nodes, with a message size of $|M|/G$.
Packing the data and flag leads to a $1 < \eta < 2$ factor increase in the message
size. Each step requires a single exchange between peers and thus, the
communication time is:
\begin{align}
  T_{\mathrm{RD}} = \log_2(N)\alpha_{\mathrm{inter}} 
  + \frac{N-1}{N}\left( \frac{\eta |M|}{G\,\beta_{\mathrm{inter}}} \right)
\end{align}
\noindent \textbf{All-gather Phase.}
Finally, the results are aggregated within each node by NCCL's all-gather
(Ring), modeled as:
\begin{align}
  T_{\mathrm{AG}} = (G-1)\alpha_{\mathrm{intra}} 
  + \frac{G-1}{G}\left( \frac{|M|}{\beta_{\mathrm{intra}}} \right)
\end{align}
\noindent \textbf{Total Communication Time.} Combining the three phases,
we get the total communication time for NVRAR as:
\begin{align}
  T_{\mathrm{NVRAR}} 
  &= 2(G-1)\alpha_{\mathrm{intra}} + \log_2(N)\alpha_{\mathrm{inter}} \nonumber\\
  &\quad + \frac{|M|}{G}\!\left[ \frac{2(G-1)}{\beta_{\mathrm{intra}}}
  + \frac{(N-1)\eta}{N\,\beta_{\mathrm{inter}}} \right]
  \label{eq:nvrar_time}
\end{align}
%
%
For small messages, the communication time is latency-dominated and the
bandwidth terms are negligible. In this regime, NVRAR scales
as $\mathcal{O}(\log_2 N)$ with the number of nodes, whereas Ring
all-reduce~\eqref{eq:ring_allreduce} scales linearly. NVRAR therefore matches
the logarithmic scaling of Tree all-reduce~\eqref{eq:tree_allreduce}, but has a
lower inter-node latency coefficient because each recursive-doubling step
requires only a single peer exchange. 


\section{Results and Discussion}
\label{sec:results}

\begin{figure*}[t]
    \centering
    \includegraphics[width=0.33\textwidth]{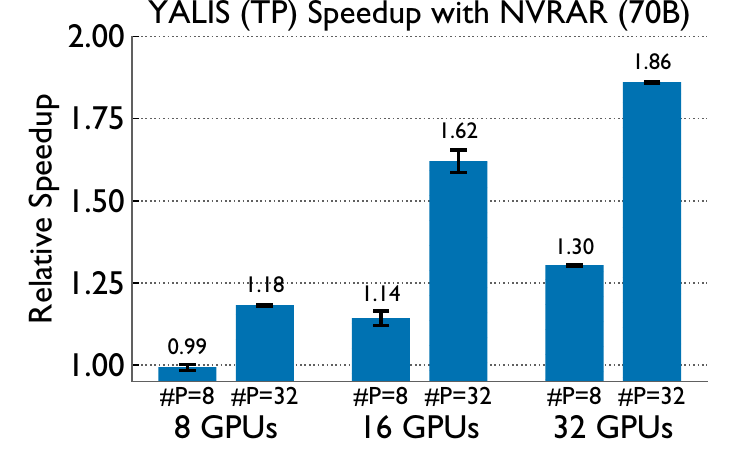}
    \includegraphics[width=0.33\textwidth]{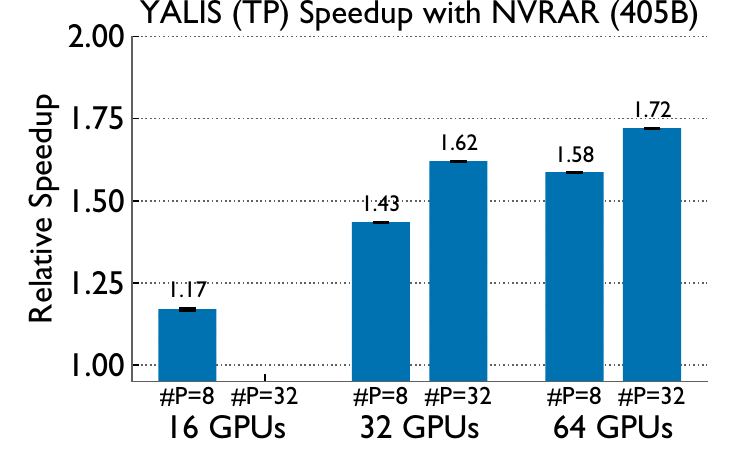}
    \includegraphics[width=0.33\textwidth]{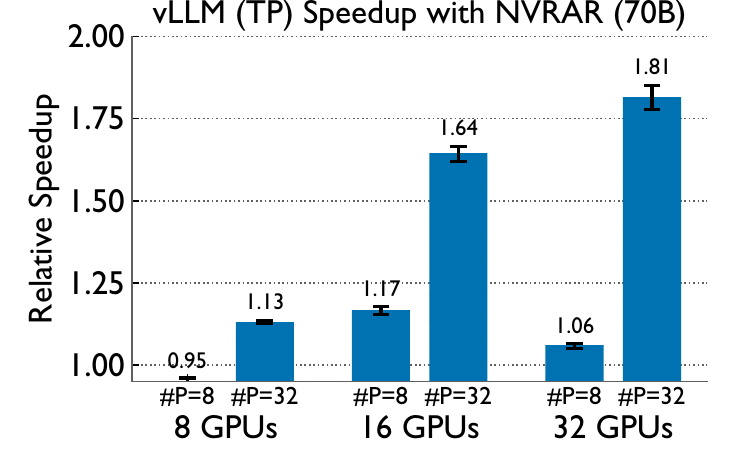}

    \caption{
      Relative speedup of YALIS (TP) and vLLM (TP) using NVRAR over NCCL
      all-reduce for the decode-heavy workload on Perlmutter,
      across different models and NumPrompts (\#P).
    }
    \label{fig:compare_pm}
\end{figure*}

This section presents a detailed performance evaluation of NVRAR against NCCL,
both as an independent collective primitive and within the context of
tensor-parallel inference workloads.

\vspace{+0.08in}
\noindent \textbf{Additional Setup Details.} We use Perlmutter and Vista
(Table~\ref{tab:machines}) for our evaluations. To isolate collective
performance, we run a microbenchmark that executes NCCL all-reduce and NVRAR,
each within a CUDA Graph, for 100 consecutive iterations. It replays the
captured graph 1,000 times (200 warm-up iterations), and we report the average all-reduce
time per collective call per iteration. CUDA Graphs help mimic inference
workloads more accurately. We use NCCL 2.27.3 and PyTorch 2.8 for all
experiments. 

For end-to-end evaluation, we integrate NVRAR into YALIS and vLLM. We run the
decode-heavy workload (Table~\ref{tab:workloads}) on both engines and compare TP
performance using NVRAR against NCCL all-reduce, reporting the relative speedup
in end-to-end batch time. We also evaluate a realistic workload trace with
1,000 prompts sampled from BurstGPT~\citep{wang2025burstgpt}, using vLLM's
benchmarking framework~\citep{vllm-benchmarking}. The trace contains a mix of
prefill- and decode-heavy requests
(Appendix~\ref{sec:appendix_burstgpt_trace_details}). For this setting, we
report the output throughput of NVRAR-based TP, NCCL-based TP, and HP
deployments. We use vLLM V1 for HP here because we do not observe the hangs
encountered in batched inference evaluations.

\subsection{Comparing NVRAR and NCCL All-reduce} 
Figure~\ref{fig:scaling_nccl_nvrar_pm} (left) reports the all-reduce microbenchmark
performance for 256\,KB and 1024\,KB messages on Perlmutter. NVRAR (orange line)
scales linearly with GPU count on a logarithmic X-axis, consistent with our
theoretical model~\eqref{eq:nvrar_time}. For 1024\,KB messages on Perlmutter,
NCCL (blue line) selects the Tree algorithm (LL protocol) at all GPU counts and
scales logarithmically as well. This setting allows us to directly compare the
two algorithms and attribute NVRAR's better performance to its lower latency
coefficients, consistent with our model. For 256\,KB messages, however, NCCL
switches from the Ring to the Tree algorithm beyond 16 GPUs, which complicates a
direct theoretical comparison. Empirically, NVRAR outperforms NCCL at most GPU
counts for both message sizes. We observe similar trends on Vista
(Appendix~\ref{sec:appendix_scaling}).

Figure~\ref{fig:scaling_nccl_nvrar_pm} (middle and right) also presents the relative
speedup of NVRAR over NCCL all-reduce for a range of message sizes and GPU
counts. On Perlmutter (middle), NVRAR is slower than NCCL for 64\,KB and 128\,KB
messages, partly due to kernel launch overheads from its three-phase design. We
also find that the microbenchmark can slightly underestimate speedup compared to
real-world workloads with interleaved communication and computation
(Appendix~\ref{sec:appendix_microbenchmark_endtoend}). Between 256\,KB and
1\,MB, NVRAR achieves significant speedups over NCCL
(1.06$\times$--1.92$\times$). On Vista (right), we observe even higher speedups.
Beyond four GPUs, for 64\,KB and 128\,KB messages, NVRAR outperforms NCCL by
1.08$\times$--1.70$\times$. Between 256\,KB and 1\,MB, NVRAR can achieve up to
3.5$\times$ lower latency than NCCL. We attribute these higher speedups to the
architecture of Vista, where each node has one GPU. As a result, NVRAR only
executes the inter-node recursive-doubling phase and has lower kernel launch
overheads.

To ensure that the reported speedups are not an artifact of suboptimal algorithm
selection in NCCL, we fix the all-reduce algorithm to Tree and Ring individually
and confirm that NVRAR's gains persist (Appendix~\ref{sec:appendix_scaling}). We
further verify that results hold across NCCL versions by comparing NCCL 2.27.3
(used in our evaluation) against NCCL 2.28.9 (shipped with the latest PyTorch 2.11).
Both versions perform similarly, and NVRAR's speedups hold against both
(Appendix~\ref{sec:appendix_new_v_old_nccl}).

\subsection{Improving Multi-node TP Inference}
We now present the end-to-end performance evaluation when using NVRAR 
with YALIS and vLLM in multi-node inference settings.

\subsubsection{Batched Inference Performance.}
Figure~\ref{fig:compare_pm} (left and middle) reports the relative speedup of
YALIS (TP) with NVRAR over NCCL all-reduce for the 70B and 405B models on
Perlmutter under the decode-heavy workload. For the 70B model (left), NVRAR
accelerates YALIS (TP) by 1.3$\times$ for NumPrompts(\#P)=8 on 32 GPUs. For this
configuration, the all-reduce message size is 128\,KB, where our standalone
microbenchmark reports slowdowns with NVRAR
(Figure~\ref{fig:scaling_nccl_nvrar_pm}). We attribute this to the 
microbenchmark launching all-reduces back-to-back without interleaved
computation, unlike real workloads. As a result, NVRAR's deferred peer
synchronization (Section~\ref{subsec:seqnum_sync}) triggers immediately at the start
of the next all-reduce, without any computation to hide it. When we run the
microbenchmark with interleaved computation, the trends match end-to-end results
(Appendix~\ref{sec:appendix_microbenchmark_endtoend}). For \#P=32 on 32 GPUs
(message size 512\,KB), NVRAR-based TP achieves a 1.86$\times$ speedup over NCCL-based TP.
For the 405B model (middle plot), speedups range from 1.17$\times$--1.72$\times$,
aided by more favorable message sizes (256\,KB and 1024\,KB) compared to the 70B
model. We observe similar speedups on Vista
(Appendix~\ref{sec:appendix_nvrar_performance}). With vLLM (TP), NVRAR delivers up to
1.81$\times$ speedup for the 70B model (Figure~\ref{fig:compare_pm} right),
demonstrating that the gains generalize beyond YALIS.

\subsubsection{Performance Breakdown.}
\begin{figure}[h]
  \centering
  \includegraphics[width=\columnwidth]{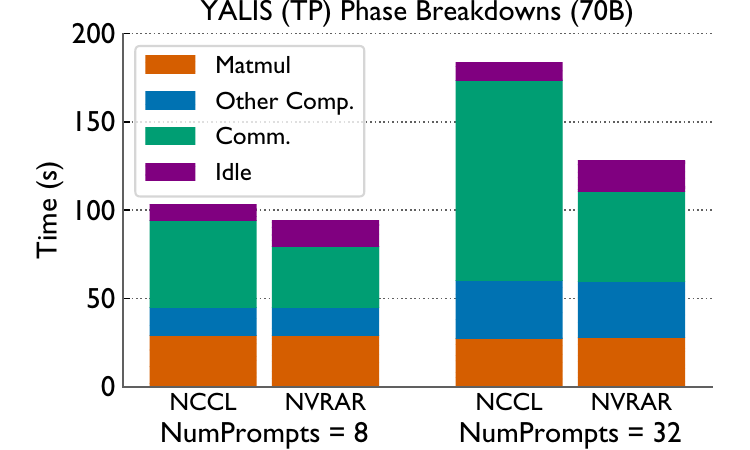}
  \caption{Per-phase time breakdown of YALIS (TP) using NVRAR and NCCL all-reduce for the decode-heavy workload on 16 GPUs of Perlmutter.}
  \label{fig:breakdown_yalis_nvrar_nccl}
\end{figure}

To better understand the performance gains observed with NVRAR, we analyze the
breakdown of end-to-end time for YALIS (TP) using NVRAR and NCCL all-reduce on
16 GPUs of Perlmutter for the 70B model in Figure~\ref{fig:breakdown_yalis_nvrar_nccl}.
For both \#P values, NVRAR leads to a lower communication time than NCCL all-reduce.
The decrease is more pronounced for \#P=32 compared to \#P=8, due to the more
favorable message size (512\,KB vs. 128\,KB). Idle time is marginally higher
for NVRAR, but not enough to offset the overall performance gains. We plan to
investigate and reduce idle time in future versions of NVRAR.

\subsubsection{Trace-based Performance Evaluation.}
\begin{figure}[h]
  \centering
  \includegraphics[width=\columnwidth]{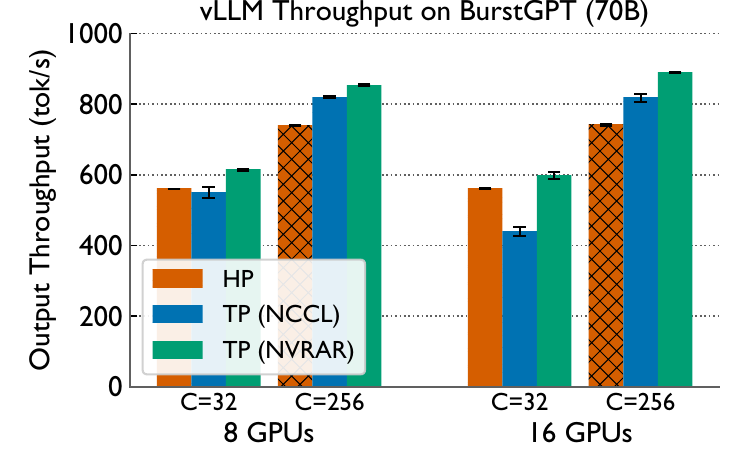}
  \caption{Output throughput on BurstGPT trace when serving the 70B model on
  Perlmutter with vLLM. We compare NCCL-based TP, NVRAR-based TP, and HP
  under two maximum request concurrency settings ($C$=32 and $C$=256).}
  \label{fig:trace_throughput}
  \vspace{-0.2cm}
\end{figure}
Finally, we evaluate NVRAR with vLLM in a serving setup, with a trace sampled
from BurstGPT. Figure~\ref{fig:trace_throughput} reports the output throughput
of NVRAR-based TP, NCCL-based TP, and HP deployments under two maximum request
concurrency settings ($C$=32 and 256).
Compared to NCCL-based TP, NVRAR delivers 1.12$\times$--1.36$\times$ higher
throughput at $C$=32. At $C$=256, the improvement decreases to
1.04$\times$--1.09$\times$, because NVRAR primarily benefits small messages
(128\,KB--4\,MB) and higher concurrency pushes the decode message sizes
towards the upper bound of this range. Compared to HP, NVRAR-based TP
achieves 1.15$\times$--1.20$\times$ higher throughput at $C$=256.
We observe smaller gains at $C$=32 in this case. This partly stems from the
mixed prefill-decode batching in vLLM. At lower concurrency, prefills are
dispersed, increasing mixed prefill-decode batches and all-reduce message size
per batch. At higher concurrency, prefills finish earlier as more
requests are processed in parallel, leading to more decode-only batches, where
NVRAR performs best. Overall, NVRAR-based TP outperforms NCCL-based TP by
1.04$\times$--1.36$\times$ across both settings, and outperforms the best
NCCL-based deployment by 1.04$\times$--1.10$\times$, validating its usefulness
for real-world workloads. For decode-heavy traces, the gains are more pronounced
(Appendix~\ref{sec:appendix_decode_heavy_trace}).

\subsubsection{Applicability to MoE Models.}
\begin{figure}[h]
  \centering
  \includegraphics[width=\columnwidth]{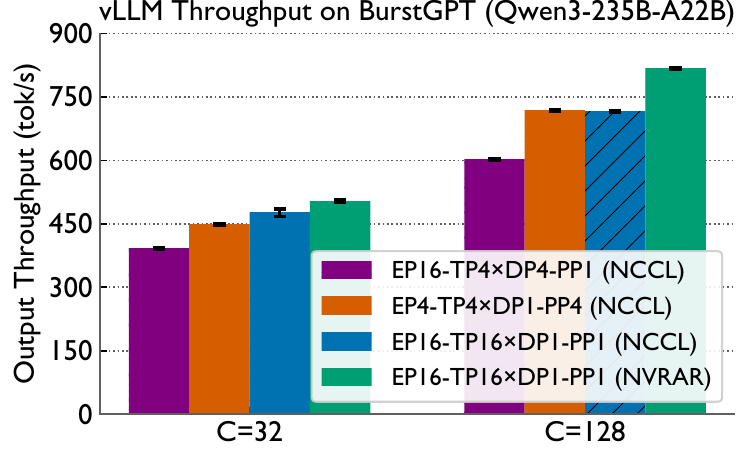}
  \caption{Output throughput on BurstGPT trace when serving Qwen3-235B-A22B on
  16 GPUs of Perlmutter with vLLM. We compare expert-parallel deployments: EP
  partitions the MoE layers, TP$\times$DP partitions the non-MoE layers, and PP
  partitions the model end-to-end. All configurations use NCCL except the last,
  which uses NVRAR. We evaluate two maximum request concurrency settings ($C$=32 and $C$=128).}
  \label{fig:moe}
\end{figure}
Mixture-of-experts (MoE) models increasingly combine expert parallelism (EP) in
the MoE layers with TP in the non-MoE layers. Since NVRAR targets TP all-reduce
communication, it is orthogonal to EP optimizations and can accelerate such MoE
deployments. To validate this, we evaluate vLLM with NVRAR on a multi-node
deployment of Qwen3-235B-A22B on 16 GPUs of Perlmutter and the BurstGPT trace.
Figure~\ref{fig:moe} compares four different parallelism configurations, all
using NCCL except the last, which uses NVRAR for the TP all-reduce. The
TP16-EP16 configuration with NVRAR achieves the highest throughput at both
concurrency settings. It achieves up to $\sim$1.14$\times$ higher throughput
than the best NCCL-based configuration at $C$=128. These results confirm that
NVRAR generalizes beyond dense models and provides meaningful gains in MoE
deployments where TP remains a critical communication bottleneck.

Our evaluation demonstrates that NVRAR achieves strong performance improvements
over NCCL all-reduce for small message sizes, and when integrated into inference
engines, it leads to significant gains in multi-node TP performance.

\section{Related Work}
\label{sec:rw}
\noindent \textbf{Model Parallel Performance Studies.} Several prior works
investigate model-parallel schemes for inference~\citep{xu2025characterizing,
zhang2025ladder, su2025seesaw, alvarez2025tensorparallelism, zhu2025nanoflow,
spector2025wholegpu}, in tandem with other system optimizations. However, these
evaluations are often limited to single-node or two-node settings and smaller
models. vLLM benchmarked Llama 3.1 405B performance~\citep{vllm_llama31_2024} on
up to 16 GPUs (two nodes), comparing InfiniBand and non-InfiniBand networks,
exposing weak TP performance on the latter. In contrast, our work presents a
detailed systematic performance study for large models in large multi-node
settings (up to 32 nodes), which is increasingly relevant. We study
and identify scaling bottlenecks for both TP and PP. For decode-bound workloads,
we further characterize PP's limited ability to reduce computation time and
attribute TP's communication overhead to sub-optimal NCCL all-reduce performance in the
small message regime.

\vspace{+0.08in}
\noindent \textbf{Collective Communication Optimization.} Recent work explores
collective communication optimizations to improve distributed inference. 
A popular approach is to overlap communication with computation.
ISO~\citep{xiao2024iso} achieves overlap in prefill, but not in decode, where
our approach is most beneficial. Ladder-residual~\citep{zhang2025ladder}
achieves overlap through model-architecture changes, which we avoid.
StragglAR~\citep{devraj2025accelerating} proposes a new all-reduce algorithm to
reduce stragglers in bandwidth-bound regimes, whereas NVRAR
focuses on latency-bound regimes. Other low-latency all-reduce optimizations, 
such as those in vLLM~\citep{woosuk2023vllm} and TensorRT-LLM~\citep{tensorrt_llm},
target NVLink-connected domains. NVRAR targets scale-out networks and is
complementary to these approaches --- its intra-node phases can be replaced with
NVLink-optimized variants. Other recent works~\citep{Zhang2025CometFC,
aimuyo2025flashdmoe, zhu2025megascaleinfer} explore NVSHMEM-based collective
optimizations for all-to-all communication in EP deployments, which are
orthogonal to NVRAR's optimizations for TP all-reduce. Finally, several
works~\citep{ueno2019hierarchical, jiang20202dhra} employ hierarchical
communication patterns for large bandwidth-bound messages typical of training
workloads. In contrast, we focus on inference workloads and target the
latency-sensitive small message regime.
\section{Conclusion}
\label{sec:conc}
In this work, we conduct a detailed performance study of model-parallelism
schemes for multi-node LLM inference workloads. We compare the performance
of Tensor Parallelism (TP) and Hybrid Parallelism (TP+PP) across
batched inference workloads and identify the scaling bottlenecks for each
strategy. Focusing on workloads that favor TP, we observe severe communication
bottlenecks arising from sub-optimal NCCL all-reduce performance for small
message sizes. Motivated by this, we develop NVRAR, a hierarchical
all-reduce implementation built using NVSHMEM. We make several key optimizations
in NVRAR to reduce latencies for small-message all-reduce operations. Compared
to NCCL, NVRAR achieves 1.06$\times$--1.92$\times$ lower latency on Slingshot
and 1.14$\times$--3.57$\times$ on InfiniBand for 256\,KB--2\,MB messages.
Integrated into YALIS and vLLM, NVRAR significantly improves multi-node TP
inference for large dense and MoE models on realistic workloads.

\begin{acks}
This work was performed in part under the auspices of the U.S. Department of
Energy by Lawrence Livermore National Laboratory under Contract
DE-AC52-07NA27344 (LLNL-CONF-2013350).

This research used resources of the National Energy Research Scientific
Computing Center (NERSC), a U.S. Department of Energy Office of Science User
Facility, operated under Contract No. DE-AC02-05CH11231 using NERSC award
DDR-ERCAP0034262 and ALCC-ERCAP0034775. This research is supported by the
National Artificial Intelligence Research Resource (NAIRR) Pilot and used the
Delta advanced computing and data resource which is supported by the NSF (award
NSF-OAC 2005572) and the State of Illinois, and the Vista supercomputing
resource at the Texas Advanced Computing Center (TACC) at The University of
Texas at Austin. The authors acknowledge the University of Maryland
supercomputing resources made available for conducting the research reported in
this paper. This work was supported by a grant from the Swiss National
Supercomputing Centre (CSCS) under project ID lp98 on Alps.


\end{acks}

\bibliography{./bib/cite,./bib/pssg}
\bibliographystyle{ACM-Reference-Format}

\clearpage
\appendix
\setcounter{equation}{0}

\section{Extended Performance Study Results}
\label{sec:appendix_perfstudy}
\begin{figure}[h]
  \centering
  \includegraphics[width=\columnwidth]{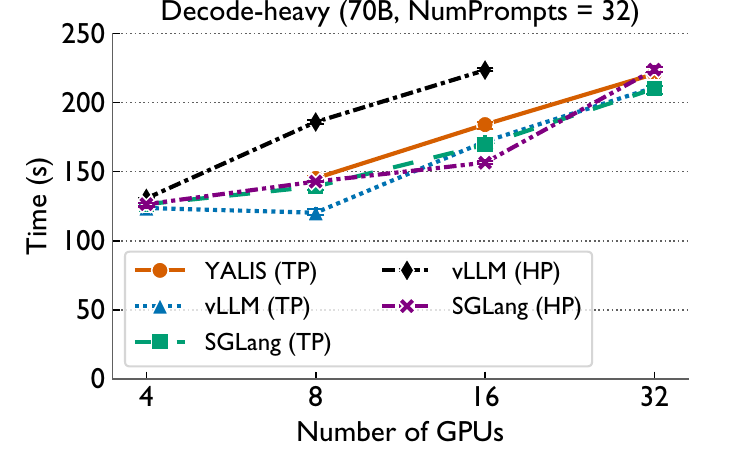}
  \\[4mm]
  \includegraphics[width=\columnwidth]{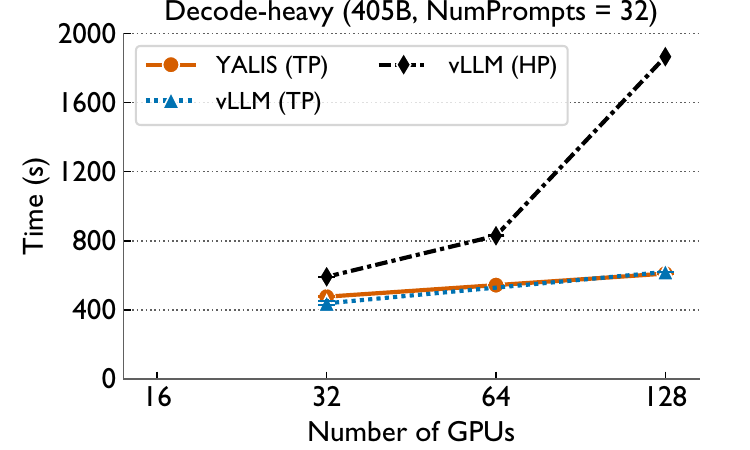}
  \caption{Strong scaling performance of different inference engines on
  Perlmutter for the Llama 3.1 70B (top) and 405B (bottom) models, for the
  decode-heavy workload with NumPrompts = 32. The Y-axis denotes the time to
  completion for a batch of prompts in seconds and the X-axis denotes the number
  of GPUs.}
  \label{fig:appendix_strong_scaling_70b_405b}
\end{figure}
Figure~\ref{fig:appendix_strong_scaling_70b_405b} presents the strong scaling
performance of different inference engines on Perlmutter for the Llama 3.1 70B
and 405B models, for the decode-heavy workload with NumPrompts = 32. We observe
that vLLM V0 (HP) (black line) scales poorly for both the 70B and 405B models.
Both TP and HP scale poorly, with total time increasing as the number of GPUs increases
for the 70B model (top). Unlike vLLM V0 (HP), SGLang (HP) has runtimes closer to
the TP configurations for the 70B model, but it still scales poorly. This closely
matches our observation in the main text.

\section{Reconciling Microbenchmark and End-to-End Performance}
\label{sec:appendix_microbenchmark_endtoend}
\begin{figure}[h]
  \centering
  \includegraphics[width=\columnwidth]{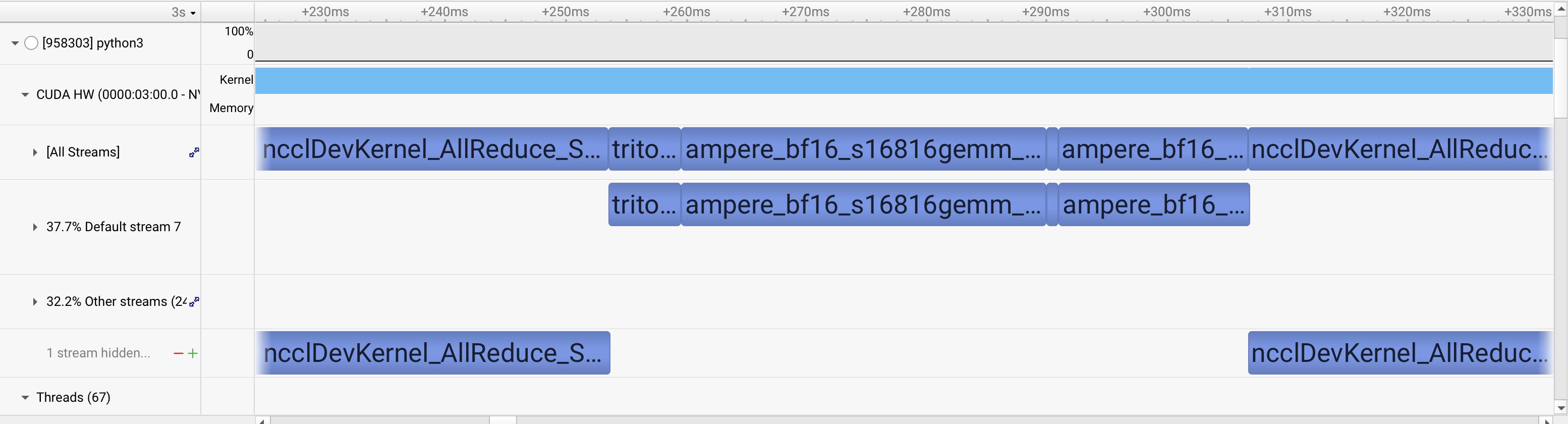}
  \caption{NSYS profile snapshots of the NVRAR kernel in the microbenchmark setup (left) and YALIS (right).}
  \label{fig:yalis_workload}
\end{figure}

\begin{figure}[h]
  \centering
  \includegraphics[width=\columnwidth]{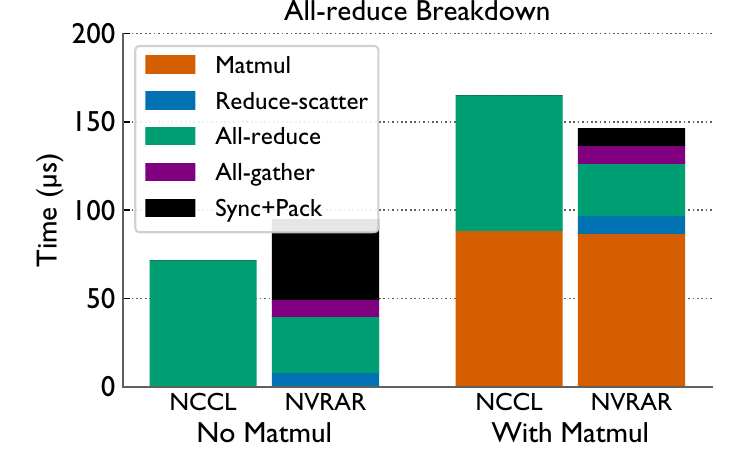}
  \caption{All-reduce time breakdown of NVRAR and NCCL for a 128\,KB message
  on Perlmutter (16 GPUs), with and without interleaved matmul computation.}
  \label{fig:reconcile_bench}
\end{figure}

In this section, we resolve the discrepancy observed in the speedup of NVRAR
over NCCL all-reduce in the standalone benchmark compared to the end-to-end
workload. For the standalone benchmark, we launch communication kernels
back-to-back, in a single CUDA Graph, without any computation work in between.
This is not representative of the real-world use cases where collective
communication operations are interleaved with computation.
Figure~\ref{fig:yalis_workload} illustrates an example YALIS trace where the
all-reduce operation is followed by several matmul and compute kernels. We
account for this difference by running synthetic matrix multiplications
between the all-reduce calls in our standalone benchmark.

Figure~\ref{fig:reconcile_bench} breaks down the 128\,KB all-reduce
microbenchmark on Perlmutter (16~GPUs) for NVRAR and NCCL, without and with an
interleaved representative matmul. Without matmul, NVRAR spends a large fraction
of time in synchronization. With interleaved matmuls, the synchronization costs
are hidden by the matmul computation due to the deferred peer-wise synchronization
strategy of NVRAR.

\section{Extended NVRAR Evaluation}
\begin{figure*}[t]
  \centering
  \includegraphics[width=0.33\textwidth]{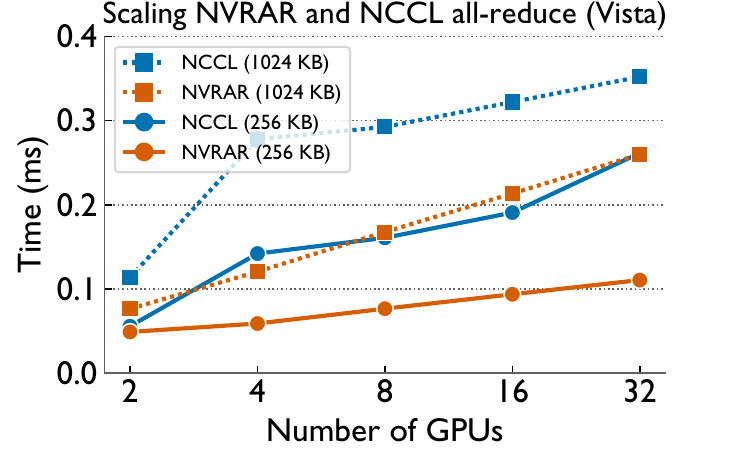}
  \includegraphics[width=0.33\textwidth]{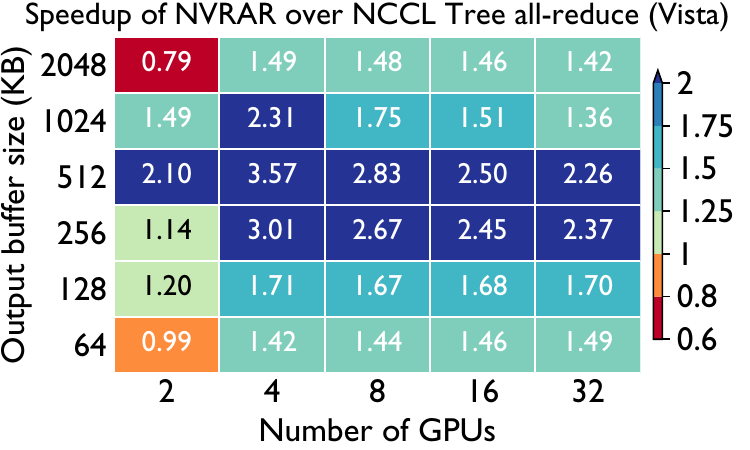}
  \includegraphics[width=0.33\textwidth]{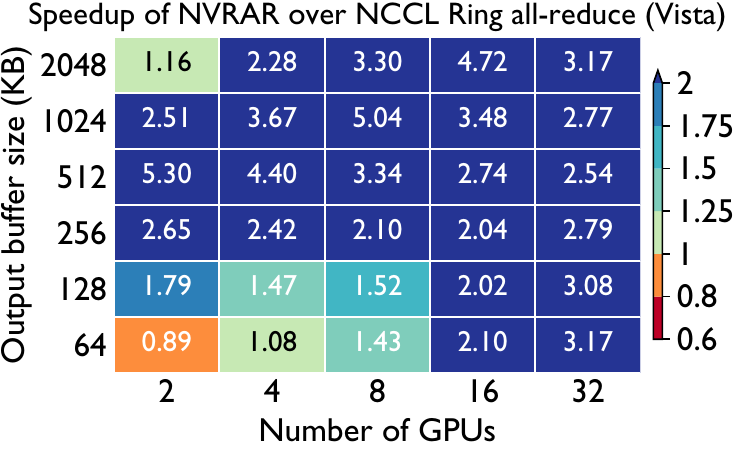}
  \caption{Performance comparison of NVRAR and NCCL all-reduce on Vista. (Left)
  Scaling of NVRAR and NCCL all-reduce for 256\,KB and 1024\,KB messages across
  GPU counts. (Middle, Right) Speedup of NVRAR over NCCL all-reduce with the NCCL
  algorithm fixed to Tree and Ring, respectively, across message sizes and GPU
  counts.}  \label{fig:vista_microbench}
\end{figure*}

This section presents an extended evaluation of NVRAR's performance, both as a
standalone collective against NCCL in microbenchmarks and as part of
tensor-parallel inference workloads in YALIS and vLLM.
\subsection{Impact of Chunk Size and Block Size on NVRAR Performance}
\label{sec:appendix_chunk_block}
We demonstrate that the hyperparameters --- block size ($B_s$) and
chunk size ($C_s$) --- have a significant impact on the performance of NVRAR.
To tune these hyperparameters, we run NVRAR with different values of $B_s$ and
$C_s$ for different message sizes and node counts.
Table~\ref{tab:chunk_block_performance} reports the performance of NVRAR with four
different hyperparameter configurations for an all-reduce message size of 1024\,KB
on 16 GPUs. We observe that the performance is more sensitive to changing the
chunk size ($C_s$) than the block size ($B_s$). This validates our design choice of
keeping these hyperparameters tunable. We also observe that chunking large
messages within each block improves performance, with smaller chunk sizes
outperforming some larger chunk sizes. We leave heuristic-based
hyperparameter tuning to future work.
\begin{table}[h]
  \caption{Different NVRAR hyperparameter configurations for an all-reduce message size of 1024\,KB on 16 GPUs.}
  \label{tab:chunk_block_performance}
  \centering
  \begin{tabular}{lcc}
    \toprule
    \textbf{Block Size ($B_s$)} & \textbf{Chunk Size ($C_s$)} & \textbf{Time (ms)} \\
    \midrule
    32 & 32768 & 0.1522 ms \\
    32 & 4096 & 0.2271 ms \\
    8 & 16384 & 0.1891 ms \\
    8 & 131072 & 0.1655 ms \\
    \midrule
    \end{tabular}
\end{table}

\subsection{Phase-Wise NVRAR Breakdown}
Figure~\ref{fig:reconcile_bench} also presents the breakdown of the time spent in
each phase of NVRAR. For the 128\,KB message size, we observe that most of
the time is spent in the inter-node recursive-doubling phase. This is expected
as the communication happens on the slower Slingshot/InfiniBand network, as
compared to the intra-node phases that happen on the faster NVLink network.

\subsection{Comparing NVRAR and NCCL All-reduce}
\label{sec:appendix_scaling}
In this section, we present additional microbenchmark results on Vista and
studies that isolate the impact of NCCL's algorithm selection and library
version on the observed speedups.

\subsubsection{Scaling Results on Vista.}
Figure~\ref{fig:vista_microbench} (left) reports the all-reduce microbenchmark
performance for 256\,KB and 1024\,KB messages on Vista, complementing the
Perlmutter results in the main text (Figure~\ref{fig:scaling_nccl_nvrar_pm}).
NVRAR scales similarly on both platforms and outperforms NCCL across all GPU
counts.

\subsubsection{NCCL Algorithm Selection.}
To ensure that NVRAR's performance gains are not an artifact of NCCL's automatic
algorithm selection, we force NCCL to use the Tree and Ring algorithms
individually and compare against NVRAR. Figure~\ref{fig:vista_microbench}
(middle, right) reports the results on Vista. NVRAR remains faster than NCCL
across the GPU range and message sizes for both algorithms, confirming that
the speedups reported in the main text are not an artifact of NCCL's algorithm
selection.

\subsubsection{Comparison Across NCCL Versions.}
\label{sec:appendix_new_v_old_nccl}
\begin{figure}[h]
  \centering
  \includegraphics[width=\columnwidth]{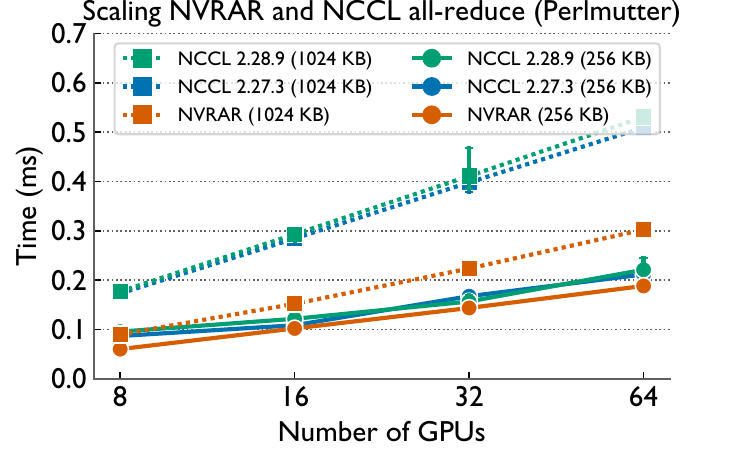}
  \caption{All-reduce scaling of NVRAR against two NCCL versions (2.27.3 and
  2.28.9) on Perlmutter for 256\,KB and 1024\,KB messages.}
  \label{fig:nccl_versions_scaling}
\end{figure}
The microbenchmark results reported in the main text use NCCL 2.27.3.
Figure~\ref{fig:nccl_versions_scaling} compares its all-reduce performance
against NCCL 2.28.9 (shipped with the latest PyTorch 2.11) on Perlmutter, alongside NVRAR.
The two NCCL versions track each other closely across all GPU counts and both
message sizes, and NVRAR retains its speedup over both. This confirms that the
speedups reported in the main text are not limited to the specific NCCL version
we evaluated against. While newer NCCL versions have several performance
improvements, most of the all-reduce optimizations are targeted towards
homogeneous NVLink-connected systems (intra-node or multi-node NVLink systems)
and are orthogonal to NVRAR's optimizations for heterogeneous networks.

\subsection{Multi-node Inference Evaluation}
\label{sec:appendix_nvrar_performance}
In this section, we present the extended performance evaluation 
when using NVRAR with YALIS and vLLM in multi-node inference settings to 
complement the results in Section~\ref{sec:results}.

\subsubsection{Batched Inference Performance on Vista.}
\label{sec:appendix_nvrar_performance_vista}
\begin{figure}[h]
  \centering
  \includegraphics[width=\columnwidth]{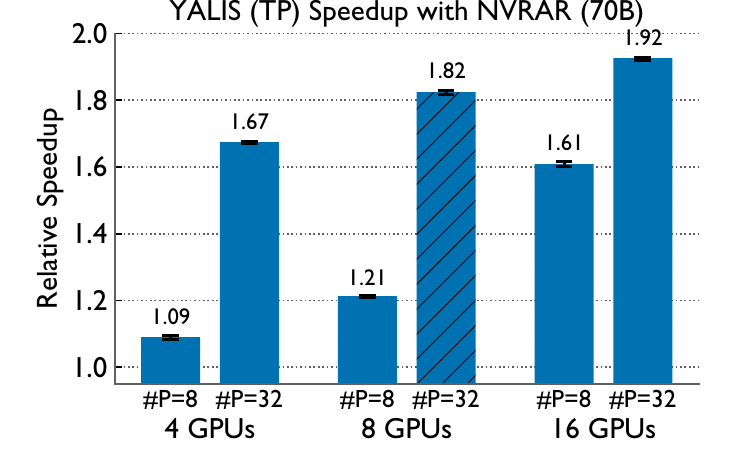}
  \caption{Relative speedup of YALIS (TP) using NVRAR all-reduce over YALIS (TP) using NCCL all-reduce for the decode-heavy workload on Vista.}
  \label{fig:compare_nvrar_vista}
\end{figure}

Figure~\ref{fig:compare_nvrar_vista} reports the relative speedup of YALIS (TP)
using NVRAR all-reduce over YALIS (TP) using NCCL all-reduce for the
decode-heavy workload on Vista. We observe that for the 70B model, NVRAR
achieves a speedup of 1.92$\times$ for NumPrompts=32 on 16 GPUs. This is
consistent with our observation on Perlmutter in the main text, validating that
NVRAR provides performance benefits on both Slingshot and InfiniBand networks.

\subsubsection{BurstGPT Trace Details.}
\label{sec:appendix_burstgpt_trace_details}
\begin{figure}[h]
  \centering
  \includegraphics[width=\columnwidth]{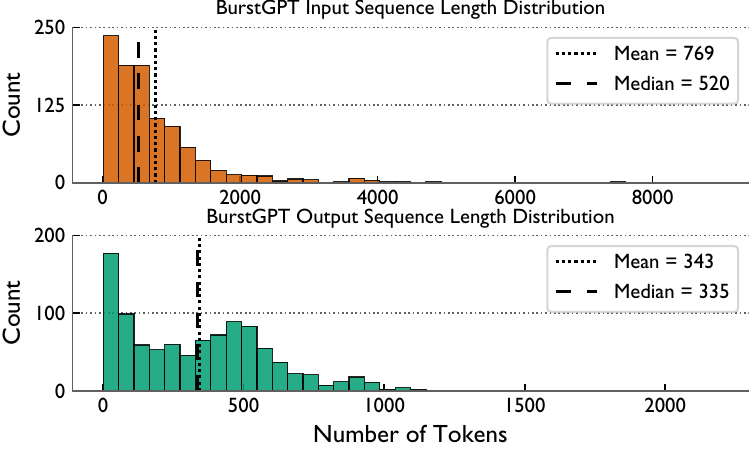}
  \caption{Input sequence length and output sequence length distribution for the BurstGPT trace (1,000 prompts).}
  \label{fig:trace_dist}
\end{figure}

Figure~\ref{fig:trace_dist} presents input sequence length and output sequence
length distributions for the BurstGPT trace used in the trace-based performance
evaluation (Figure~\ref{fig:trace_throughput}). Table~\ref{tab:trace_details}
provides the details of the vLLM benchmark configuration.

\begin{table}[h]
  \caption{vLLM benchmark settings for trace-based performance evaluation.}
  \label{tab:trace_details}
  \centering
  \begin{tabular}{ll}
    \toprule
    \textbf{Setting} & \textbf{Value} \\
    \midrule
    Concurrency & 32, 256 \\
    Number of Prompts & 1,000 \\
    Configured Request Trace & 10 requests/second \\
    Burstiness & 2.0 (Gamma distribution) \\
    \bottomrule
    \end{tabular}
\end{table}

\subsubsection{Evaluation on Decode-heavy Trace.}
\label{sec:appendix_decode_heavy_trace}
\begin{figure}[h]
  \centering
  \includegraphics[width=\columnwidth]{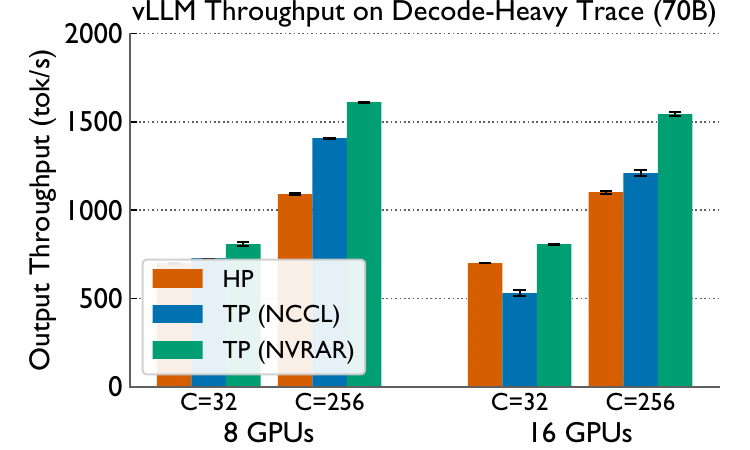}
  \caption{Output throughput on a randomly generated decode-heavy trace when serving
  the 70B model on Perlmutter with vLLM. We compare NCCL-based TP, NVRAR-based TP, and HP deployments under two
  maximum request concurrency settings ($C$=32 and $C$=256).}
  \label{fig:trace_throughput_dh}
  \vspace{-0.2cm}
\end{figure}
In addition to BurstGPT, we evaluate on a randomly generated decode-heavy trace
with mean input and output sequence lengths of 1024 and 4096, respectively. We use the
same vLLM benchmark settings as in Table~\ref{tab:trace_details}.

Figure~\ref{fig:trace_throughput_dh} reports the output throughput of
NVRAR-based TP, NCCL-based TP, and HP for the decode-heavy trace. We observe
that in this case, NVRAR-based TP outperforms NCCL-based TP by
1.11$\times$--1.52$\times$ and HP by 1.16$\times$--1.48$\times$ across these
settings, and outperforms the best NCCL-based deployment by
1.11$\times$--1.28$\times$. These are higher than the speedups observed on the
BurstGPT trace (Figure~\ref{fig:trace_throughput}). This is expected, as
NVRAR is more beneficial in the decode phase, with small to medium message sizes.
This is consistent with our batched inference performance evaluations.

\end{document}